\documentclass{emulateapj}
\usepackage{amstext,amsmath,natbib}

\shorttitle{Ionized Lines in 4U~1916$-$05}
\shortauthors{JUETT AND CHAKRABARTY}
\slugcomment{Accepted for publication in the Astrophysical Journal}

\begin{document}

\newcommand{\source}{4U~1916$-$05}

\title{Detection of Highly Ionized Metal Absorption Lines in the
Ultracompact X-ray Dipper 4U 1916$-$05}

\author{Adrienne~M.~Juett\altaffilmark{1} and
Deepto~Chakrabarty\altaffilmark{2}}

\altaffiltext{1}{Department of Astronomy, University
of Virginia, Charlottesville, VA 22903; ajuett@virginia.edu} 

\altaffiltext{2}{Department of Physics and Kavli Institute for
Astrophysics and Space Research, Massachusetts Institute of
Technology, Cambridge, MA 02139; deepto@space.mit.edu}

\begin{abstract}
We present the high-resolution {\em Chandra X-ray Observatory}
persistent (non-dip) spectrum of \source\/ which revealed narrow
absorption lines from hydrogenic neon, magnesium, silicon, and sulfur,
in addition to the previous identified hydrogenic and helium-like iron
absorption lines.  This makes \source\/ only the second of the
classical X-ray dipper systems to show narrow absorption lines from
elements other than iron.  We propose two possible explanations for
the small measured line widths ($\lesssim500$--2000~km~s$^{-1}$),
compared to the expected Keplerian velocities ($> 1000$~km~s$^{-1}$)
of the accretion disk in this 50-min orbital period system, and lack
of wavelength shifts ($\lesssim 250$~km~s$^{-1}$). First, the ionized
absorber may be stationary.  Alternatively, the line properties may
measure the relative size of the emission region.  From this
hypothesis, we find that the emission region is constrained to be
$\lesssim$0.25 times the radial extent of the absorber.  Our results
also imply that the ionized absorber spans a range of ionization
parameters.
\end{abstract}


\section{Introduction}
It was anticipated that high-resolution spectra of low-mass X-ray
binaries (LMXBs) would reveal a wealth of line features which could be
used to better understand accretion disk structure, but such features
have been largely absent in most {\em Chandra X-ray Observatory} and
{\em XMM-Newton} spectra of LMXBs.  One exception is the class of
X-ray dippers.  X-ray dippers show intensity dips, although not
necessarily eclipses, on the orbital period and have inclination
angles in the range $60^{\circ}$$<$$i$$<$$80^{\circ}$.  The dipping
behavior is caused by absorption of the continuum by the accretion
disk or its atmosphere.

The best studied of the X-ray dippers, the neutron star (NS) binary,
EXO~0748$-$676 ($P_{\rm orb}$$=$3.8~hr), showed emission lines from
the hydrogenic and helium-like species of nitrogen, oxygen, neon,
magnesium, and silicon in its {\em XMM} and {\em Chandra} spectra
\citep{ckb+01,jsm03}.  The {\em XMM} spectrum of the X-ray dipper
MXB~1659$-$298 ($P_{\rm orb}$$=$7.1~hr) showed narrow absorption lines
from highly ionized oxygen, neon, and iron, and a broad Fe-$K$
emission line \citep{sop+01}.  {\em XMM} data from four other dippers,
X1624$-$490 ($P_{\rm orb}$$=$21.0~hr), 4U~1254$-$69 ($P_{\rm
orb}$$=$3.9~hr), 4U~1916$-$05 ($P_{\rm orb}$$=$0.83~hr), and
4U~1323$-$62 ($P_{\rm orb}$$=$2.94~hr) revealed similar narrow iron
absorption lines and some evidence for broad iron emission lines
\citep{pob+02,bp03,bpb+04,bmt+05,crd+05}.

The presence of these absorption features in X-ray dippers, but not
in other LMXBs, has led to the suggestion that the absorbing material
has a cylindrical distribution \citep[e.g.,][]{bpb+04}.  The lack of
variation in the properties of the absorption lines as a function of
orbital phase, excluding the dipping region, implies an azimuthal
symmetry \citep[see, e.g.,][and references therein]{bpb+04}.  This in
turn points to some relationship between the accretion disk and the
absorbing material.

In this paper, we present the high-resolution {\em Chandra} spectrum
of \source, the shortest orbital period X-ray dipper.  The 50-min
orbital period of 4U~1916$-$05 was discovered independently by
\citet{ws82} and \citet{wbm+82} from dips seen in the X-ray emission.
The stability of the dip period over many years led the authors to
associate it with the orbital period of the system \citep[but also
see,][]{cgb01,hch+01,rcb+02}.  The dipping behavior of \source\/ is
variable on the order of $\sim$4 days, varying from short
($\approx$10\% of orbital phase), shallow dips to long ($\approx$40\%
of orbital phase), deep dips \citep[e.g.,][]{cgb01}.  Unique among the
X-ray dippers, the 50-min orbital period of \source\/ places it in the
class of ultracompact LMXBs which require hydrogen-deficient and/or
degenerate donors \citep[e.g.,][]{jar78}.  It has been proposed that
the companion is a hydrogen-deficient but not yet degenerate star
\citep{nrj86}.  Observational evidence points to a helium-rich donor.
The X-ray burst properties of \source\/ suggest that the accreting
material is helium-rich and that the distance to \source\/ is
8.9$\pm$1.3~kpc (Galloway et al. 2005, in prep.).  A recent optical
spectrum of \source\/ shows a feature at 4540~\AA\/ which may be due
to \ion{He}{2} \citep{nj04}.

The non-dip spectrum can be fit with a power-law or power-law $+$
blackbody model in the range 0.5--10 keV.  Variations in the best-fit
spectral parameters are correlated with the broad-band luminosity and
the position of \source\/ in its color-color diagram \citep{bgb+00}.
In addition, \citet{adn+00} detected a broad (FWHM~$=$~0.7~keV)
emission feature in the {\em ASCA} spectrum at 5.9 keV with an
equivalent width of 87 eV attributed to Fe-$K$ emission.  Highly
ionized iron absorption lines were found in the {\em XMM} spectrum of
4U~1916$-$05 with marginal detections of other highly ionized features
\citep[e.g., \ion{Mg}{12}, \ion{S}{16};][]{bpb+04}.

\section{Observations and Data Reduction}
We observed \source\/ on 2004 August 07 for 50 ks with {\em Chandra}
using the High Energy Transmission Grating Spectrometer (HETGS) and
the Advanced CCD Imaging Spectrometer \citep[ACIS;][]{cdd+05}.  The
HETGS carries two transmission gratings: the Medium Energy Gratings
(MEGs) with a range of 2.5--31~\AA\/ (0.4--5.0~keV) and the High
Energy Gratings (HEGs) with a range of 1.2--15~\AA\/ (0.8--10.0~keV).
The HETGS spectra are imaged by ACIS, an array of six CCD detectors.
The HETGS/ACIS combination provides both an undispersed (zeroth order)
image and dispersed spectra from the gratings.  The various orders
overlap and are sorted using the intrinsic energy resolution of the
ACIS CCDs.  The first-order MEG (HEG) spectrum has a spectral
resolution of $\Delta\lambda=$ 0.023~\AA\/ (0.012~\AA).  To reduce
pileup, the observation used a sub-array of 512 rows, yielding a
frametime of 1.7~s.  The ``level 1'' event files were processed using
the CIAO v3.2 data analysis
package\footnote{http://asc.harvard.edu/ciao/}.  The standard CIAO
spectral reduction procedure was performed.

\begin{figure}
\epsscale{0.7}
\plotone{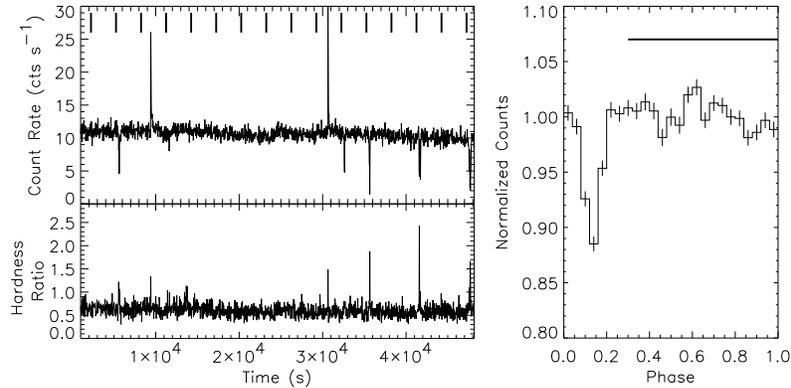}
\caption{{\em Left panel, top:} First order HEG$+$MEG 0.5--8.0~keV
count rate over the {\em Chandra} observation of \source.  The solid
marks at the top of the figure indicate the expected dip times using
the dip ephemeris of \citet{cgb01}.  {\em Left panel, bottom:} The
hardness ratio ([4--8~keV count rate]/[0.5--1.5~keV count rate])
during the observation.  As expected the hardness ratio increases
during the X-ray dips (due to low energy absorption) and X-ray bursts.
{\em Right Panel:} Normalized phase folded lightcurve of the first
order HEG$+$MEG 0.5--8.0~keV counts from \source.  The times of X-ray
bursts have been removed and the lightcurve has been corrected for
exposure effects.  The solid line from phases 0.3--1.0 indicates the
data used to produce the persistent (non-dip) spectrum.}
\label{fig:lc}
\end{figure}

The combined MEG $+$ HEG first order dispersed spectrum of \source\/
has an average count rate of 10.6$\pm$1.3~cts~s$^{-1}$.  During our
observation, \source\/ showed only mild dipping behavior as well as
two X-ray bursts (see Figure~\ref{fig:lc}).  In the analysis presented
here, we focus only on the persistent (non-dip) emission from \source.
We inspected the lightcurve at times of expected dipping, using the
dip ephemeris of \citet{cgb01}.  Of the 16 orbital cycles covered by
our observation, only six showed statistically significant dipping
behavior.  These dips were centered on a phase of $\approx0.15$ and
had width of $\approx0.15$ in phase.  To produce our persistent
(non-dip) spectrum, we excluded data from phases 0.0--0.3 for all
orbital cycles.  In addition, we excluded data during the two X-ray
bursts.  We filtered the level 2 event file, yielding a persistent
(non-dip, non-burst) exposure time of 31~ks.  We then extracted source
and background spectra for both the plus and minus first order MEG and
HEG.  Detector response files (RMFs and ARFs) were created for the
four spectra using the standard CIAO tools.

For bright sources, pileup can be a problem for CCD detectors
\citep[see, e.g.,][]{d03}.  We checked the dispersed spectra of
\source\/ and found signs of pileup in the first order MEG spectrum.
In dispersed spectra, pileup can affect only a limited wavelength
range, particularly where the effective area of the instrument is the
highest.  In our observation, pileup was present between 2--10 \AA\/
in the MEG plus and minus first orders.  No pileup was found in the
HEG spectra.

The spectral analysis was performed using ISIS \citep{hd00}.  The $+1$
and $-1$ order spectra were combined for the HEG and MEG respectively
using the ISIS function {\tt combine\_datasets}.  This function is
similar to summing the datasets, but more accurately takes into
account the different responses for each dataset.  We fit jointly the
HEG spectra over the range 1.6--11.5~\AA\/ and MEG spectra over the
range 10.0--14.0~\AA.

\section{Analysis and Results}
\begin{figure}
\epsscale{0.8}
\plotone{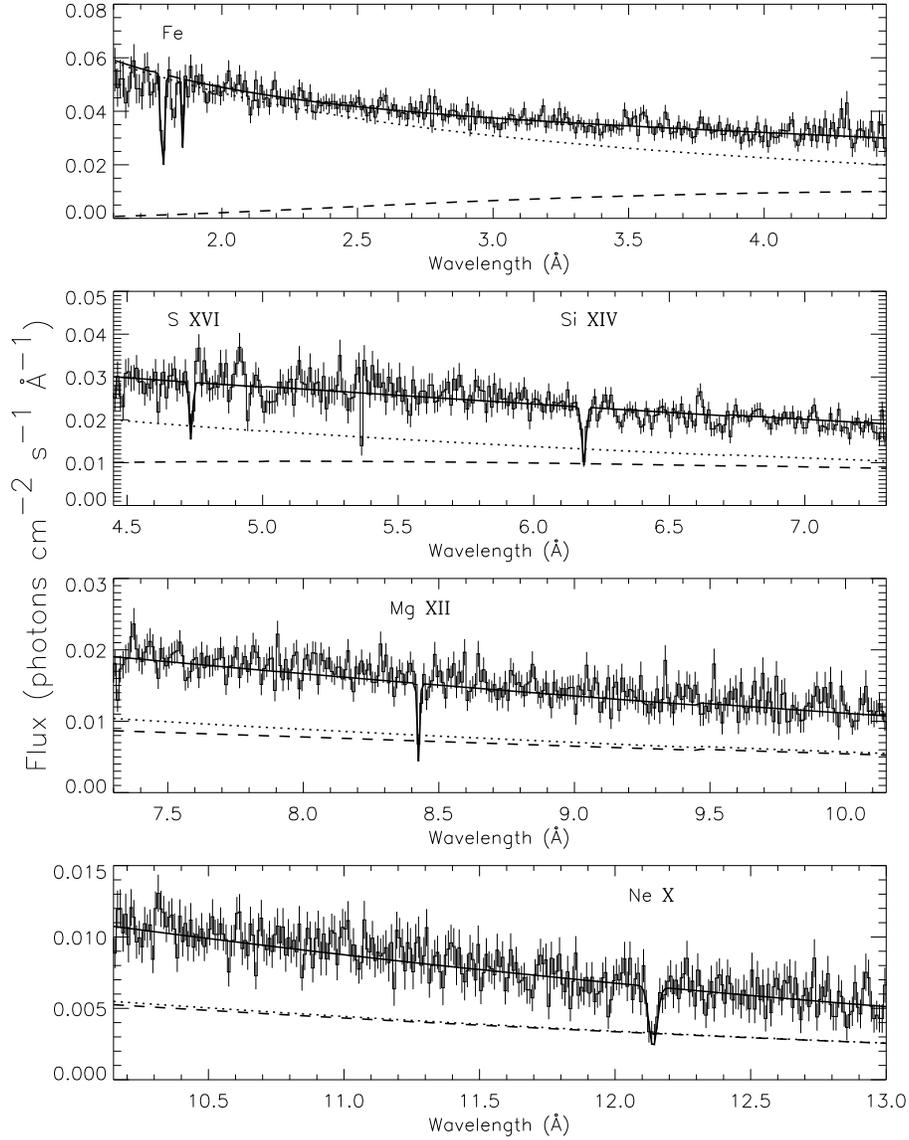}
\caption{Flux spectra and best-fit model for the combined first order
MEG$+$HEG {\em Chandra} spectrum of \source.  The data and model are
binned to 0.010~\AA.  The solid line is the full best-fit model
including both continuum and line components.  The dotted and dashed
lines show the best-fit power law and disk blackbody continuum models,
respectively.  The absorption features from neon, magnesium, silicon,
sulfur, and iron are labeled.}
\label{fig:flux}
\end{figure}

An initial inspection of the spectrum of \source\/ revealed narrow
absorption features at 12.1, 8.4, 6.2, and 4.7~\AA\/ (see
Figure~\ref{fig:flux}), attributable to hydrogenic neon, magnesium,
silicon, and sulfur.  To determine the continuum model, we fit the
spectrum ignoring regions around these lines as well as the Fe-$K$
region at 1.8~\AA.  We tested three different continuum models: power
law, power law $+$ blackbody, and power law $+$ disk blackbody, all
with interstellar absorption as described by the {\tt tbabs} model
using the abundances of \citet{wam00}.  Both the power law $+$
blackbody and power law $+$ disk blackbody models give better fits to
the data than the power law alone (see Table~\ref{tab:cont}).  We take
the power law $+$ disk blackbody model as the continuum model,
although we note that the choice of the power law $+$ blackbody model
would make little difference to our results.  The absorbed 0.5--10~keV
flux from \source\/ was $8.0\times 10^{-10}$~erg~cm$^{-2}$~s$^{-1}$.

Our best-fit model is significantly different from that found by
\citet{bpb+04}.  But as noted by \citet{bpb+04}, the limited energy
range covered by {\em XMM} and {\em Chandra} make it difficult to
determine a unique solution to the continuum fit.  One interesting
difference between our {\em Chandra} data and the {\em XMM} data, is
the presence of an edge at 0.98~keV.  \citet{bpb+04} found that both
the EPIC and RGS data were best-fit when an edge, with depth $\tau =
0.11\pm0.03$, was included.  In contrast, the addition of an edge at
0.98~keV does not improve the chi-squared value of the continuum fit
for the {\em Chandra} data.  We find an upper limit of $\tau < 0.08$
for the edge, barely consistent with the {\em XMM} result.  The
0.5--10~keV luminosity of \source\/ was $9.0 \times
10^{36}$~erg~s$^{-1}$, using a distance of 8.9 kpc, during the {\em
Chandra} observation.  This is $\approx$2 times greater than found
during the {\em XMM} observation \citep{bpb+04}.

\begin{figure}
\epsscale{1.0}
\plotone{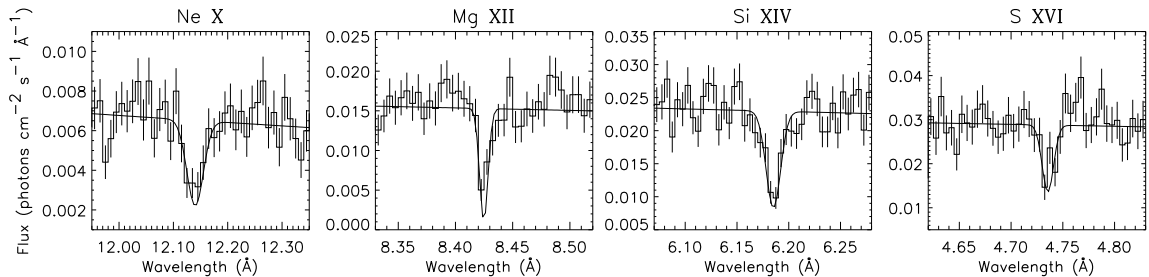}
\caption{Flux spectra and best-fit models for the \ion{Ne}{10},
\ion{Mg}{12}, \ion{Si}{14}, and \ion{S}{16} line regions.  Note that
the \ion{Ne}{10} line is covered by the lower resolution MEG, while
the other lines are covered by the HEG.  The line-like feature at
$\approx$11.97~\AA\/ in the \ion{Ne}{10} region is more narrow than
the instrument resolution and is likely an instrumental artifact.}
\label{fig:lowz}
\end{figure}

\begin{figure}
\epsscale{0.7}
\plotone{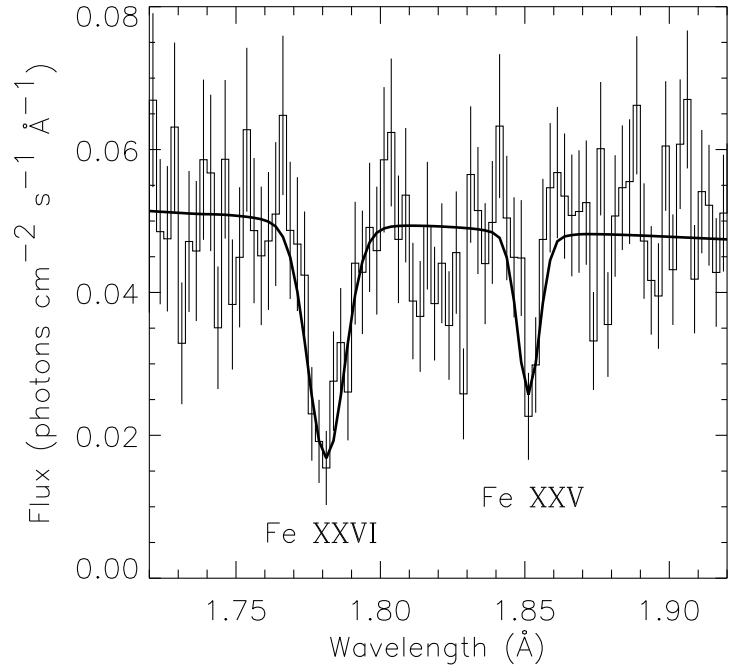}
\caption{Flux spectrum and best-fit model for the iron line region.
The \ion{Fe}{25} line is found at 1.85~\AA\/ and the \ion{Fe}{26} line
is found at 1.78~\AA.}
\label{fig:fe}
\end{figure}

With the best-fit continuum model fixed, we fit the lines from
\ion{Ne}{10}, \ion{Mg}{12}, \ion{Si}{14}, \ion{S}{16}, \ion{Fe}{25},
and \ion{Fe}{26} with Gaussian models to determine the line positions,
widths, and fluxes.  In addition, we determined the upper limits to
the flux from the helium-like ions of neon, magnesium, silicon, and
sulfur.  For the upper limit measurements, we fixed the wavelength to
those given in \citet{bn02} and we fixed the line width $\sigma =
0.005$~\AA.  We marginally detected the helium-like ion of sulfur.
The neon lines were covered by the MEG spectrum, while the rest of the
lines were located in the HEG spectral region.  The best-fit
parameters are given in Table~\ref{tab:line} and the spectra and
best-fit models are shown in Figures~\ref{fig:lowz} and \ref{fig:fe}.

The line positions are consistent with the rest frame wavelengths of
hydrogenic neon, magnesium, silicon, and sulfur, and hydrogenic and
helium-like iron.  This suggests that there is no bulk motion along
our line of sight associated with the absorbing material.  In
addition, the measured line widths are comparable to, or less than,
the instrument resolution (0.023~\AA\/ for MEG, 0.012~\AA\/ for HEG).
Therefore, direct measurements of the velocity widths, or temperature,
of the absorbing material are not possible with these data.  The
derived temperature limits range from $<4 \times 10^{8}$~K for
\ion{Mg}{12}, to $1.2\pm1.1 \times 10^{10}$~K for \ion{Fe}{26}.

From the best fit equivalent widths (EW), we can estimate the column
density ($N_{\rm Z}$) of the ions using the relationship between EW,
$N_{\rm Z}$, and the transition oscillator strength $f_{ij}$
\begin{equation}
\frac{W_{\lambda}}{\lambda} = \frac{\pi e^2}{m_e c^2} N_{\rm Z}
\lambda f_{ij} = 8.85 \times 10^{-13} \, N_{\rm Z} \lambda f_{ij},
\end{equation}
where $W_{\lambda}$ is the EW in wavelength units and $\lambda$, the
line wavelength, is given in cm units \citep{s78}.  The equation above
is true only when a line is unsaturated, on the linear part of the
curve of growth.  For saturated lines, the exact relationship between
EW and $N_{\rm Z}$ becomes a more complicated function, and includes a
dependence on the velocity of the ions.  We used the oscillator
strengths given by \citet{vvf96} for the hydrogenic lines, and those
of \citet{bn02} for the helium-like lines.  Our results are given in
Table~\ref{tab:line}.  If the lines are saturated, the values
represent a lower limit to the column densities for each ion.

\begin{figure}
\epsscale{0.7}
\plotone{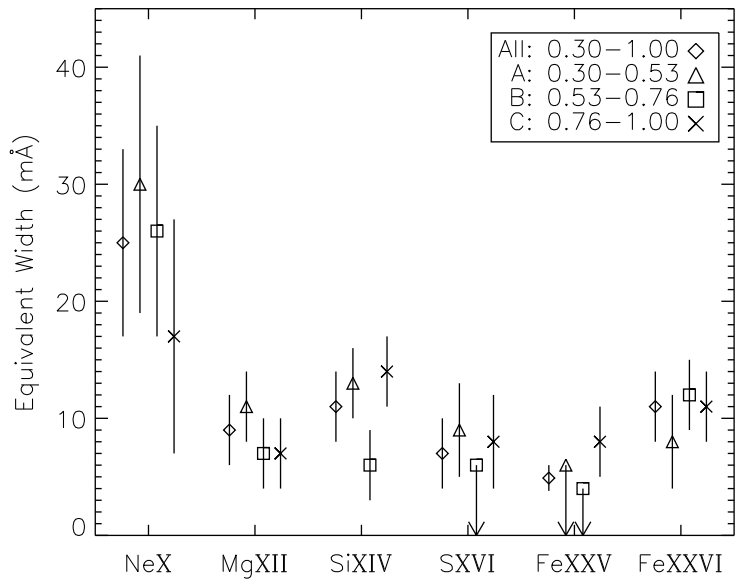}
\caption{Comparison of the absorption line equivalent widths as a
function of orbital phase.  We plot the EWs for the full dataset
(diamond), in addition to three phase groups: group A covering phases
0.30--0.53 (triangle), group B covering phases 0.53--0.76 (square),
and group C covering phases 0.76--1.00 (x).  No significant variation
of the line properties are found.}
\label{fig:comp}
\end{figure}

Comparing our results with those of \citet{bpb+04}, we find that the
iron line EWs are barely consistent within the errors.  Our
\ion{Fe}{26} EW is at the upper limit of the {\em XMM} result, while
our \ion{Fe}{25} EW is at the lower limit.  It is possible that the
greater luminosity of \source\/ during the {\em Chandra} observation
has caused a real difference in the \ion{Fe}{25} and \ion{Fe}{26}
column densities, compared to the lower luminosity {\em XMM} data, but
the difference is not significant enough to confirm this.  The iron
line widths found in the {\em Chandra} data are lower than the upper
limits found in the {\em XMM} data owing to the higher resolution of
the HETG compared to the EPIC pn.  Our EW measurements for \ion{S}{16}
and \ion{Ne}{10} are consistent with the values reported for the {\em
XMM} data, but our \ion{Mg}{12} EW is significantly below the result
of \citet{bpb+04}.  We note however that \citet{bpb+04} concluded that
their \ion{Mg}{12} detection was only marginal, therefore the
difference between the {\em Chandra} and {\em XMM} results may not
indicate a real change between the two observations.  As in the other
X-ray dipper observations, we found no variation in the line
parameters as a function of orbital phase (see Figure~\ref{fig:comp}).

\section{Discussion}
Our {\em Chandra}/HETGS observation of \source\/ revealed narrow,
unresolved absorption lines in the persistent emission, attributable
to hydrogenic neon, magnesium, silicon, and sulfur, in addition to the
previous identified hydrogenic and helium-like iron absorption lines.
This makes \source\/ only the second of the classical X-ray dipper
systems to show narrow absorption lines from mid-$Z$ elements
\citep[and references therein]{bpb+04}.

The properties of these lines is of particular interest for
understanding the emission and absorption processes.  With a 50~min
orbital period, \source\/ is the shortest period X-ray dipper, and as
such should have the smallest accretion disk.  If the lines are
associated with the accretion disk, they should have widths comparable
to the expected Keplerian velocities in the disk.  Assuming a 1.4--2.0
$M_{\odot}$ primary and a 0.1--0.15 $M_{\odot}$ secondary \citep[see,
e.g.][]{nrj86}, we can estimate the outer disk velocity assuming that
the disk fills 70\% of the primary Roche lobe.  We find that the outer
disk velocity should be in the range of 1000--1300~km~s$^{-1}$, while
the companion velocity should be 740--840~km~s$^{-1}$.  In all cases,
the measured line widths are consistent with, or below, the instrument
resolution.  For \ion{Mg}{12} and \ion{Fe}{25}, the upper limits on
the line widths are less than the estimated outer disk velocity.
Additionally, we measure no shift in the wavelength of the lines,
ruling out a strong outflow.  The \ion{S}{16} line shows a slight
shift ($+$0.004~\AA) but given that this is the least significant of
our line detections and the possibility of systematic errors, we do not
feel it is a significant result.

We suggest two possible explanations for these results.  First, the
absorber could be static, producing the narrow lines and lack of
wavelengths shifts.  This however seems unlikely given the large
quantities of angular momentum present in the system.  Instead, we
propose that the line properties are a measure of the emission
properties, particularly the extent of the emission.  In this case, we
assume that the ionized absorber is associated with either the
accretion disk or its atmosphere, as suggested by the implied
cylindrical geometry \cite[see e.g.,][]{bpb+04}.  The rotation of the
absorbing material is only measurable when it has a component along
the line of sight to the emitter.  If the bulk velocity is
perpendicular to the line of sight, no velocity effects will be found
in the absorption lines (excluding thermal or turbulent motions which
we take as small compared to the Keplerian velocities in the disk).
From the limits on the linewidths, we can calculate the maximum
allowable radial extent of the emitter.  We find that the X-ray
emission region is $\lesssim 32000$~km, assuming that the absorber is
located at the outer edge of the accretion disk.  This represents the
maximum upper limit on the extent.  We note that if the absorber is
located at a smaller disk radius, the upper limit will be smaller,
roughly 1/4 of the absorber radius.

Assuming that photoionization is the dominant ionization process for
the absorbing material, we can estimate the ionization parameter, $\xi
= L/n_{e} r^{2}$ \citep{tts69}, from the ratio of the helium-like to
hydrogenic column densities for each atomic species.  We assume that
the helium-like and hydrogenic ions are present at the same,
single-valued ionization parameter.  This is a simple approximation to
the expected true state of the system in which the ions exist in
different relative proportions over a range of ionization parameters.

We used
XSTAR\footnote{http://heasarc.gsfc.nasa.gov/docs/software/xstar/xstar.html}
to estimate the ion abundances as a function of $\xi$ for an optically
thin plasma, with constant number density and solar abundances.  We
note that while the companion in \source\/ is expected to be a
hydrogen-deficient star, and solar abundances would therefore not
apply, we found that varying the abundances made no significant
difference in the helium-like to hydrogenic column density ratios for
any element.  The only XSTAR input that made a substantial difference
in the inferred ionization parameters was the shape of the input
spectrum.  We tried three models, a power law with $\alpha = -1$, a
10~keV bremsstrahlung, and a user defined model based on the continuum
emission of \source.

We compared the XSTAR results with the column density ratios for neon,
magnesium, silicon, sulfur, and iron as given in Table~\ref{tab:line}.
We note that these values are only valid if the lines are unsaturated.
The limits for neon and magnesium are too high to provide any
constraint on the ionization parameter.  For silicon, we find $\log
\xi \gtrsim$1.7--2.0 and for sulfur, $\log \xi = 1.5$--3.3, given the
error in the measurement and the variation in $\xi$ with spectral
model (highest values for the power law model, lower for the
bremsstrahlung and user defined models).  The most constraining result
comes of course from the iron ratio, where we find $\log \xi =$
3.50$\pm$0.13 for the bremsstrahlung model, 3.73$\pm$0.13 for the
power law model, and 3.02$\pm$0.12 for the user defined model.  This
rough estimate would suggest that the iron absorption is found in a
material with higher ionization than the lower $Z$ elements.  Our
result is consistent with modeling of the ionized absorber in
MXB~1658$-$298 which required a distribution of ionization parameters
\citep{tpb+05}.  This is of course expected since in a material
ionized enough to contain helium-like and hydrogenic iron, the lower
$Z$ elements would be completely stripped.

\acknowledgements{We thank Jonathan Gelbord, Kazunori Ishibashi, and
Michael Nowak for useful discussions.  We would also like to thank Tim
Kallman for his help with XSTAR.  We also thank the anonymous referee
whose comments improved the paper.  Partial support for this work was
provided by NASA through Chandra award number GO4-5054X.}

\clearpage
\begin{deluxetable}{lcccccc}
\tabletypesize{\footnotesize} 
\tablewidth{0pt} 
\setlength{\tabcolsep}{0.1in}
\tablecaption{Best-Fit Continuum Spectral Parameters\tablenotemark{a}}
\tablehead{\colhead{Model} & \colhead{$N_{\rm H}$ ($10^{21}$ cm$^{-2}$)} 
  & \colhead{$\Gamma$} & \colhead{$A_{1}$\tablenotemark{b}} & 
  \colhead{$kT$ (keV)} & \colhead{Norm\tablenotemark{c}} & 
  \colhead{$\chi^2_{\nu}/ \nu$} } 
\startdata
PL          &  6.9$\pm$0.2 &  1.519$\pm$0.019 &  12.2$\pm$0.3 & 
     \nodata       & \nodata     &  1.123/1187 \\
PL$+$BB     &  5.6$\pm$0.7 &  1.31$\pm$0.11   &  8.5$\pm$1.7  & 
     0.54$\pm$0.04 &  70$\pm$30  &  1.114/1185 \\
PL$+$DISKBB &  5.4$\pm$0.7 &  1.0$\pm$0.4     &  5$\pm$3      & 
     0.91$\pm$0.11 &  15$\pm$5   &  1.113/1185 \\
\enddata
\tablenotetext{a}{All errors quoted at the 90\%-confidence level.}
\tablenotetext{b}{Power-law normalization in units of $10^{-2}$
photons cm$^{-2}$ s$^{-1}$ keV$^{-1}$.}  
\tablenotetext{c}{Thermal component normalization.  For the blackbody
model, norm $= R^2_{\rm km}/D^2_{\rm 10 \, kpc}$; for the disk
blackbody model, norm $= (R_{\rm in,km}/D_{\rm 10 \, kpc})^2
\cos\theta$, where $\theta$ is the inclination angle of the disk.}
\label{tab:cont}
\end{deluxetable}

\begin{deluxetable}{lccccc}
\tabletypesize{\footnotesize} 
\tablewidth{0pt} 
\setlength{\tabcolsep}{0.1in}
\tablecaption{Best-Fit Line Parameters\tablenotemark{a}}
\tablehead{\colhead{Ion} & \colhead{Wavelength (\AA)} & 
  \colhead{FWHM (\AA)} & \colhead{FWHM (km~s$^{-1}$)} & \colhead{EW (m\AA)} & 
  \colhead{$N_{\rm Z}$\tablenotemark{b} ($10^{16}$~cm$^{-2}$)}} 
\startdata
\ion{Ne}{10} & 12.140$\pm$0.006  & 0.035$\pm$0.019 & 870$\pm$470  & 
     $-25\pm$8    & 4.6$\pm$1.5 \\
\ion{Mg}{12} & 8.425$\pm$0.002   & $<$0.014        & $<$500       & 
     $-9\pm$3     & 3.4$\pm$1.1 \\
\ion{Si}{14} & 6.185$\pm$0.002   & 0.016$\pm$0.009 & 800$\pm$460  & 
     $-11\pm$3    & 8$\pm$2 \\
\ion{S}{16}  & 4.736$\pm$0.003   & 0.012$\pm$0.009 & 750$\pm$600  & 
     $-7\pm$3     & 8$\pm$4 \\
\ion{Fe}{25} & 1.8519$\pm$0.0017 & $<$0.005        & $<$950       & 
     $-4.9\pm$1.1 & 20$\pm$5 \\
\ion{Fe}{26} & 1.7811$\pm$0.0015 & 0.012$\pm$0.005 & 1900$\pm$900 & 
     $-11\pm$3    & 90$\pm$30 \\ \tableline
\multicolumn{6}{c}{Upper Limits on Helium-like Ions} \\
\tableline
\ion{Ne}{9}  & 13.448 (fixed)    & 0.012 (fixed)   & 260 (fixed)  & 
     $> -9$       & $< 0.9$ \\
\ion{Mg}{11} & 9.170 (fixed)     & 0.012 (fixed)   & 380 (fixed)  & 
     $> -6$       & $< 1.2$ \\
\ion{Si}{13} & 6.648 (fixed)     & 0.012 (fixed)   & 530 (fixed)  & 
     $> -3$       & $< 1.1$ \\
\ion{S}{14}  & 5.039 (fixed)     & 0.012 (fixed)   & 700 (fixed)  & 
     $-4\pm$3     & 2.5$\pm$1.9 \\
\enddata
\tablenotetext{a}{All errors quoted at the 90\%-confidence level.}
\tablenotetext{b}{Assumes lines are unsaturated.  Values represent a
lower limit to the column density if lines are saturated.}
\label{tab:line}
\end{deluxetable}

\end{document}